\documentclass[aip,twocolumn,reprint]{revtex4-2}
\usepackage{graphicx}% Include figure files
\usepackage{dcolumn}% Align table columns on decimal point
\usepackage{bm}% bold math
\usepackage[utf8]{inputenc}
\usepackage[T1]{fontenc}
\usepackage{mathptmx}
\usepackage{etoolbox}
\usepackage{epstopdf}

\draft % marks overfull lines with a black rule on the right

\begin{document}

%\preprint{}

\title{Loss investigations of high frequency lithium niobate Lamb wave resonators at ultralow temperatures} %Title of paper



\author{Wenbing Jiang}
\thanks{Authors to whom correspondence should be addressed: wbjiang@siom.ac.cn; lbzhou@siom.ac.cn; and wutao@shanghaitech.edu.cn}
\affiliation{Wangzhijiang Innovation Center for Laser, Aerospace Laser Technology and System Department, Shanghai Institute of Optics and Fine Mechanics, Chinese Academy of Sciences, Shanghai, 201800, China}
\affiliation{Center of Materials Science and Optoelectronics Engineering, University of Chinese Academy of Sciences, Beijing, 100049, China}

\author{Xuankai Xu}
\affiliation{School of Information Science and Technology, ShanghaiTech University, Shanghai, 201210, China}

\author{Jiazhen Pan}
\affiliation{Purple Mountain Laboratories, Nanjing, 211111, China}

\author{Hancong Sun}
\affiliation{Purple Mountain Laboratories, Nanjing, 211111, China}

\author{Yu Guo}
\affiliation{Wangzhijiang Innovation Center for Laser, Aerospace Laser Technology and System Department, Shanghai Institute of Optics and Fine Mechanics, Chinese Academy of Sciences, Shanghai, 201800, China}
\affiliation{Center of Materials Science and Optoelectronics Engineering, University of Chinese Academy of Sciences, Beijing, 100049, China}

\author{Huabing Wang}
\affiliation{Purple Mountain Laboratories, Nanjing, 211111, China}

\author{Libing Zhou}
\thanks{Authors to whom correspondence should be addressed: wbjiang@siom.ac.cn; lbzhou@siom.ac.cn; and wutao@shanghaitech.edu.cn}
\affiliation{Wangzhijiang Innovation Center for Laser, Aerospace Laser Technology and System Department, Shanghai Institute of Optics and Fine Mechanics, Chinese Academy of Sciences, Shanghai, 201800, China}
\affiliation{Center of Materials Science and Optoelectronics Engineering, University of Chinese Academy of Sciences, Beijing, 100049, China}

\author{Tao Wu}
\thanks{Authors to whom correspondence should be addressed: wbjiang@siom.ac.cn; lbzhou@siom.ac.cn; and wutao@shanghaitech.edu.cn}
\affiliation{School of Information Science and Technology, ShanghaiTech University, Shanghai, 201210, China}
\affiliation{Shanghai Engineering Research Center of Energy Efficient and Custom AI IC, Shanghai, 201210, China}
%\email[]{}


\date{\today}

\begin{abstract}
Lamb wave resonators (LWRs) operating at ultralow temperatures serve as promising acoustic platforms for implementing microwave-optical transduction and radio frequency (RF) front-ends in aerospace communications because of the exceptional electromechanical coupling ($k^2$) and frequency scalability. However, the properties of LWRs at cryogenic temperatures have not been well understood yet. Herein, we experimentally investigate the temperature dependence of the quality factor and resonant frequency in higher order antisymmetric LWRs down to millikelvin temperatures. The high-frequency A1 and A3 mode resonators with spurious-free responses are comprehensively designed, fabricated, and characterized. The quality factors of A1 modes gradually increase upon cryogenic cooling and shows 4 times higher than the room temperature value, while A3 mode resonators exhibit a non-monotonic temperature dependence. Our findings provide new insights into loss mechanisms of cryogenic LWRs, paving the way to strong-coupling quantum acoustodynamics and next-generation satellite wireless communications.
\end{abstract}


\maketitle %\maketitle must follow title, authors, abstract and \pacs


Gigahertz acoustic phonons with low inherent loss and the shorter wavelength compared to electromagnetic waves provide a versatile landscape for coupling with the classical and quantum states in microwave and optical domain \cite{delsing2019,clerk2020hybrid,hassanien2021efficient,iyer2024coherent}. For example, strong coupling between acoustic wave resonators in the quantum regime and superconducting qubits has been achieved to construct circuit quantum acoustodynamics (cQAD) \cite{manenti2017circuit}, quantum memory \cite{hann2019hardware}, and microwave-optical transducers connecting distant quantum nodes \cite{blesin2024bidirectional,zhao2025quantum}. Moreover, the temperature sensitivity of acoustic wave resonators renders them suitable for cryogenic temperature sensors and RF filters in aerospace environments \cite{crupi2021measurement,zhang2024ultra,zheng2024temperature}. In recent decades, surface acoustic wave (SAW) resonators have been successfully employed for cQAD, aerospace sensors, energy harvesting, and acousto-optic modulation \cite{manenti2017circuit,bolgar2018quantum,ruan2024tunable,le2022noncontact,le2023evolution,xu2024unveiling}, and the ultralow-temperature properties of SAW resonators were thoroughly studied as well \cite{magnusson2015surface,manenti2016surface,yamamoto2023low,lee2024cryogenic,liu2025comparative}. 

More recently, LWRs have attracted growing interest for wideband RF filters applications and quantum transduction, owing to their superior $k^2$ relative to SAW resonators, the excellent frequency scalability, and more concentrated acoustic energy because of the suspended structure \cite{yang20194,lu2020a1,liu20227,Luo2022,Kramer2025,mirhosseini2020superconducting}. Accordingly, the low temperature behaviors of high-frequency LWRs, particularly at the millikelvin temperatures, have been demanded to elucidate the underlying loss mechanisms and guide optimizations of the device performance. Nevertheless, previous characterizations of LWRs with cryogenic cooling were restricted to the liquid nitrogen temperature range\cite{zheng2024temperature,tu2016effects,kramer2024experimental}, and short of the comprehensive temperature dependent behaviors across the entire temperature range from room temperature down to ultralow temperatures because of the poor cryogenic temperature control in the measurement setup. Recently, we have also reported the enhancement in the quality factor (\emph{Q}) of S0 mode LWRs on the Al$_{0.7}$Sc$_{0.3}$N platform at liquid helium temperatures, with superconducting NbN films as electrodes \cite{Wenzhen2025}. However, the acoustic resonant frequency is limited below 1 GHz, and the operating temperature is insufficiently low for quantum applications. To date, comprehensive investigations into the performance and loss mechanism of high-frequency LWRs at millikelvin temperatures remain lacking.

In this study, we have performed an experimental investigation on the energy loss of high-frequency LWRs at millikelvin temperatures on the thin-film lithium niobate (LN) platform. With the sophisticated design of higher order antisymmetric Lamb modes, we have successfully fabricated and characterized spurious-free A1 and A3 mode devices on the Z-cut thin-film LN. The resonant frequencies of LWRs are tailored to be larger than 4 GHz, compatible with the operating frequency range of superconducting qubits \cite{gao2025establishing}. The temperature dependence of \emph{Q}-factors for A1 and A3 mode resonators were extracted and discussed in detail, guiding the optimization direction of high-\emph{Q} LWRs for applications in quantum acoustodynamics, quantum transducers, and front-end filters in satellite wireless communications.

Figure 1(a) shows the simulated resonant frequency of A1 and A3 modes, with the longitudinal wavelength $\lambda\textsubscript{L}$ of 28 $\mu$m, for electrically short ($f\textsubscript{short}$) and open ($f\textsubscript{open}$) conditions on the top surface with tuning the LN film thickness. The resonant frequency of both modes decreases for thicker LN films due to the smaller wavenumber along the thickness propagation direction. We chose the LN film thickness of 400 and 700 nm for A1 and A3 modes, respectively, where the acoustic frequency locates around 4 and 7 GHz satisfying the requirement of coupling with superconducting qubits. Furthermore, the calculated $k^2$ with respect to $\lambda\textsubscript{L}$ is shown in Fig. 1(b), based on the formula \cite{lu2020a1}:
\begin{equation}
\emph{k}^2 = \frac{\pi^2}{8}\cdot[(\frac{\emph{f}\textsubscript{open}}{\emph{f}\textsubscript{short}})^2-1].  \label{Eq1} 
\end{equation}
We determine $\lambda\textsubscript{L}$ = 28 $\mu$m where the $k^2$'s of A1 and A3 modes attain the maximum.   

\begin{figure}
  \centering
  \includegraphics[width=\linewidth]{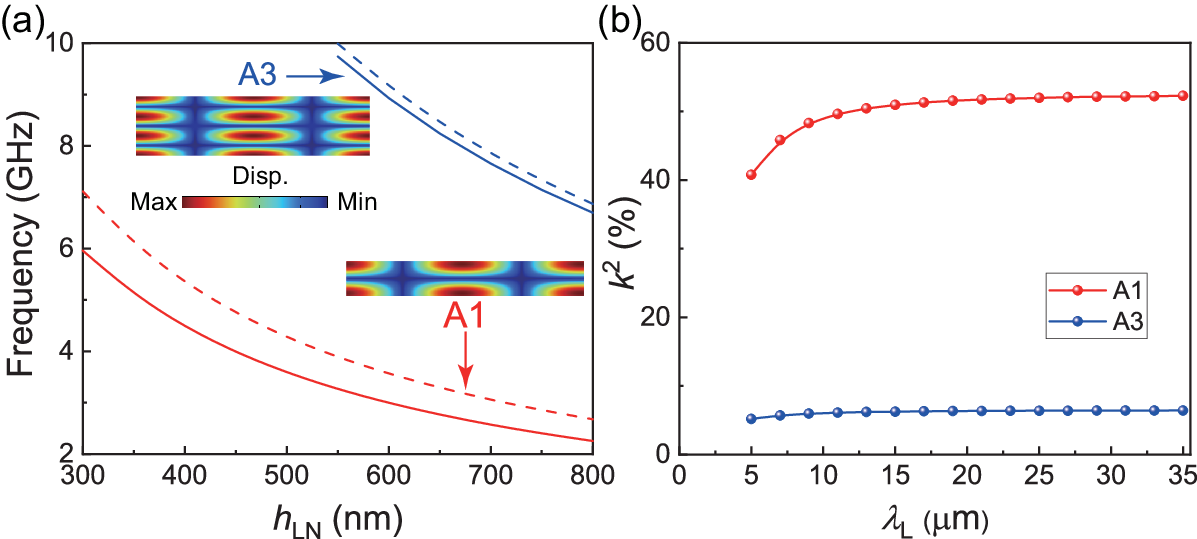}
  \caption{(a) The simulated resonant frequency of A1 and A3 modes as a function of LN film thickness, respectively. The solid and dash lines represent the resonant frequency for electrically open and short boundaries, respectively. The insets illustrate the cross-sectional view of the displacement mode shapes of A1 and A3 modes. (b) The simulated $k^2$ with respect to $\lambda\textsubscript{L}$ for A1 and A3 modes.}\label{Fig1}
\end{figure}

For the A1 mode, the excitation of the spurious modes is an inevitable issue to be considered in quantum devices and wideband RF filters. There are a variety of schemes to mitigate the spurious modes near the main A1 response \cite{zou2019transverse,yang2021lateral,tong20246}. The recessed electrode design has been adopted in our design. Fig. 2(a) shows the simulated admittance of three different designs, namely conventional electrodes and recessed electrodes with different electrode thickness (70 nm and 100 nm), respectively. There exists a pronounced in-band spurious mode at 5 GHz for the conventional electrode design. With the recessed electrode configuration (70 nm), the in-band spurious mode is remarkably suppressed and the admittance spectrum of the A1 mode becomes clean. However, if the recessed electrode thickness is further increased to 100 nm, a kink around the resonant frequency $f\textsubscript{s}$ reappears. Therefore, we adopt the recessed electrode design with the thickness of aluminum (A1) electrode, $h\textsubscript{Al}$ = 70 nm, in the actual device fabrication. Al is selected because its low density and acoustic impedance, which minimize the mass-loading effect, while its superconducting transition at low temperature allows direct examination of the influence of electrode ohmic loss on resonator dissipation. In contrast, heavier metals such as Au, Cu, and Mo would increase mass loading and acoustic damping \cite{zhang2023high}, introducing more spurious modes and degrading the device performance, as shown in Fig. 2(d).

The fabrication error of etching depths for recessed electrodes has been investigated as shown in Fig. 2(c). The shallow etching case reveals spurious modes near the anti-resonant frequency $f\textsubscript{p}$, while the deeper etching results in the spurious-free spectra. Notably, the admittance spectra near $f\textsubscript{s}$ are robust against the etching depth and there are no spurious modes observed, which means the recessed electrode scheme possesses large fabrication tolerance for grooves. Fig. 2(d) displays the simulated frequency dependence of admittance for the A3 mode with various electrode metals and $h\textsubscript{LN}$ = 700 nm. Although there is a reduction in $k^2$ for A3 against A1 modes, the A3 resonance with Al electrodes is free from spurious modes and $f\textsubscript{s}$ can reach higher frequency ranges without additional thinning films.

\begin{figure}
  \centering
  \includegraphics[width=\linewidth]{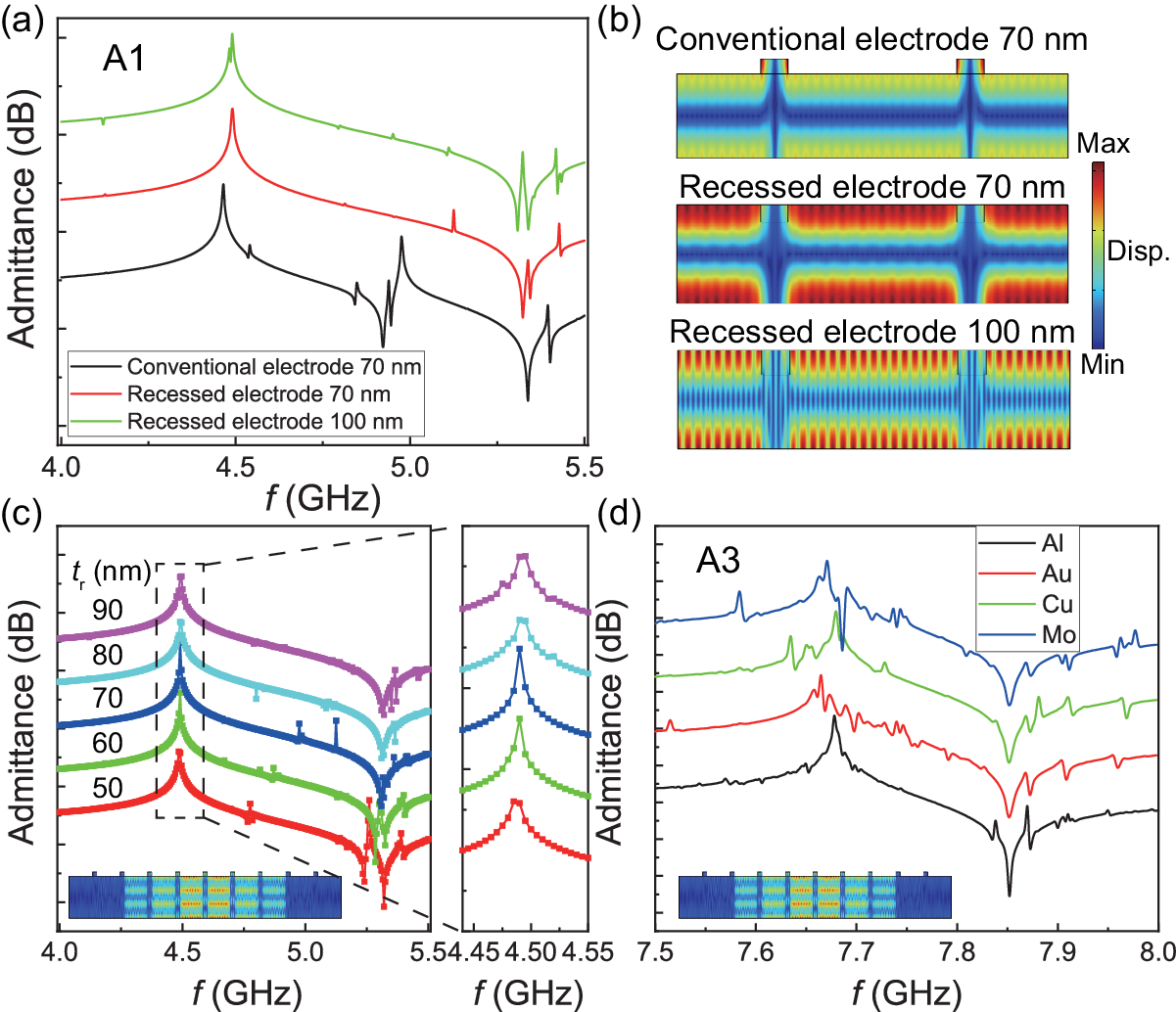}
  \caption{(a) The simulated frequency dependence of admittance for the conventional and recessed electrode design. (b) Displacement profiles at the resonant frequency for three designs. (c) Fabrication error analyses of the etching depth for recessed electrodes. (d) The admittance spectrum of the A3 mode for various electrode metals. The displacement profiles for A1 and A3 modes with Al electrodes are shown in the inset, respectively.}\label{Fig2}
\end{figure}

\begin{figure}
  \centering
  \includegraphics[width=\linewidth]{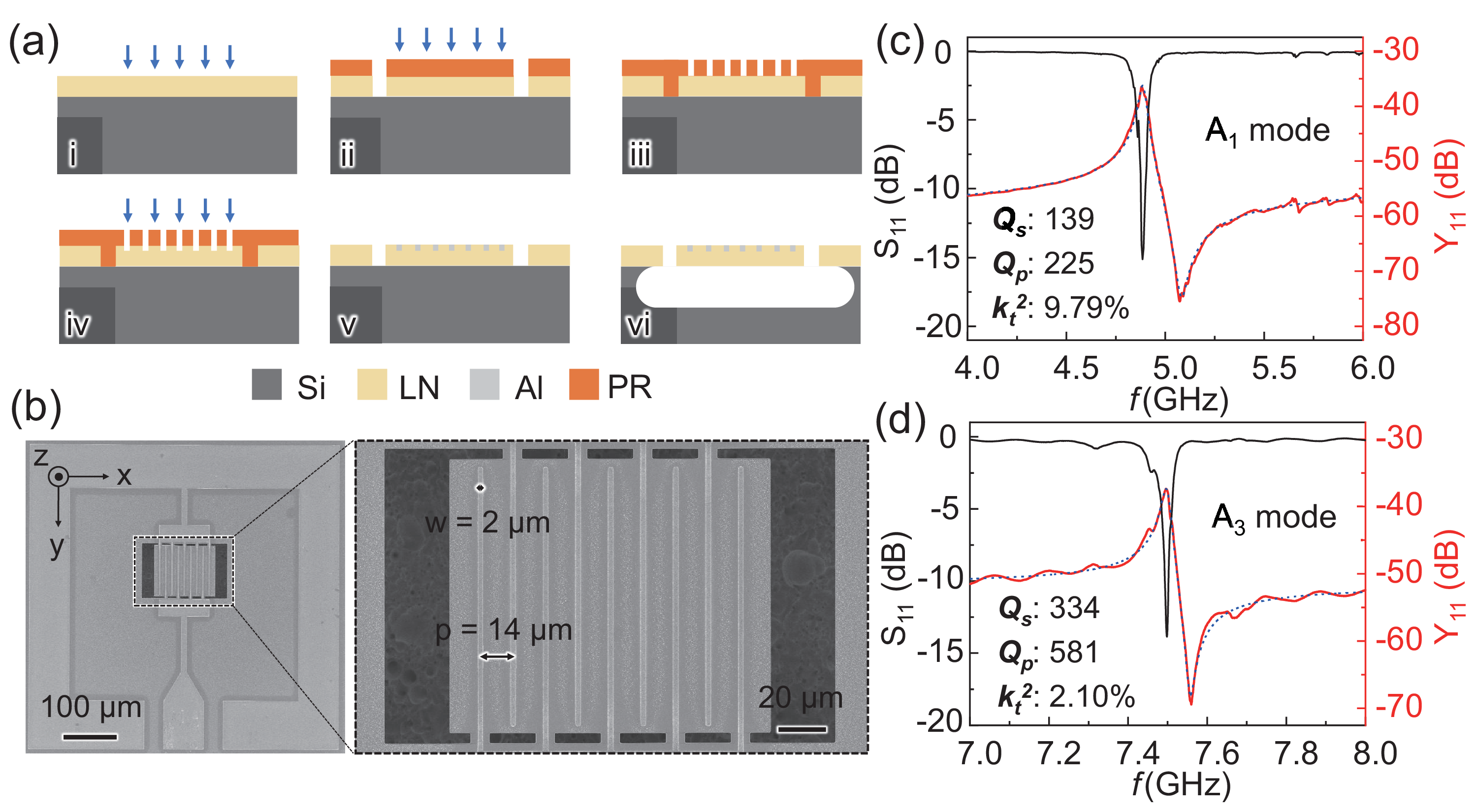}
  \caption{Fabrication process and characterization of LWRs. (a) Schematic of the six-step fabrication process: (i) Initial LN-on-Si wafer thinning via IBE; (ii) Boundary definition through photoresist patterning and IBE etching; (iii) Recessed area patterning with photoresist; (iv) Shallow etching (70 nm depth) using IBE; (v) Al electrode deposition via sputtering and lift-off; (vi) Final Si substrate release using XeF$_2$ etching. (b) SEM image of the fabricated resonator structure. Electrical characterization of $S_{11}$ parameters and $Y_{11}$ admittance for (c) A1 mode and (d) A3 mode resonators at room temperature using GSG probes.}\label{Fig3}
\end{figure}

The fabrication process of the LWRs, illustrated in Fig. 3(a), consists of six sequential steps. For the A3 mode resonator, a 700 nm LiNbO$_3$-on-Si (LN-on-Si) wafer was utilized, while the A1 mode resonator required thinning the LN film to 400 nm via ion beam etching (IBE). Next, photolithographic patterning defined the resonator boundaries, followed by IBE etching to shape the structure. To suppress spurious A1 modes, recessed regions were patterned using photoresist and shallow-etched ($\sim$ 70 nm) with an additional IBE step. This step is omitted for the A3 mode resonator, which does not exhibit spurious mode interference. Subsequently, 70 nm-thick Al electrodes were deposited by sputtering and patterned through a lift-off process. Finally, the underlying silicon substrate was selectively removed via XeF$_2$ dry etching to suspend the resonator structure. Fig. 3(b) shows a representative scanning electron microscope (SEM) image of the fabricated device. The interdigital electrodes are oriented perpendicular to the x-direction to maximize piezoelectric excitation. The design features a pitch of 14 $\mu$m, electrode widths of 2 $\mu$m, and 9 electrode fingers. 

Electrical characterization was performed at room temperature using ground-signal-ground (GSG) probes and a vector network analyzer (VNA, Keysight N5234B). The measured $S_{11}$ parameters and $Y_{11}$ admittance responses, along with Modified Butterworth-Van Dyke (MBVD) model fittings for the A1 and A3 modes \cite{lu2019accurate}, are presented in Fig. 3(c) and 3(d), respectively. The A1 mode exhibits $f\textsubscript{s}$ of 4.885 GHz and $f\textsubscript{p}$ of 5.083 GHz, yielding the series ($Q\textsubscript{s}$) and parallel ($Q\textsubscript{p}$) quality factors, $Q\textsubscript{s}$ = 139 and $Q\textsubscript{p}$ = 225, along with $k\textsubscript{t}^2$ of 9.79$\%$. In contrast, the A3 mode shows a series resonance at 7.496 GHz and an anti-resonance at 7.56 GHz, with significantly higher quality factors ($Q\textsubscript{s}$ = 334, $Q\textsubscript{p}$ = 581) but $k\textsubscript{t}^2$ of 2.10 $\%$. These measured values are in good agreement with the simulation results.

After the room temperature characterization of A1 and A3 modes, we have measured the reflection coefficient $S_{11}$ of A1 and A3 resonators at low temperatures down to millikelvin levels. The devices were wire-bonded and packaged in oxygen-free copper sample box which was anchored in the mixing chamber of the dilution refrigerator. Fig. 4(a) illustrates the diagram of low-temperature measurement setups. The input microwave signals feed through the attenuators at each temperature stage to suppress the thermal noises. Afterwards, the reflected signals from the device are routed to the circulator, groups of isolators and filters at $\sim$10 mK, the high electron mobility transistors (HEMT) at 4 K, and the low-noise RF amplifier at room temperature. Eventually, the amplified signals are acquired by the VNA. A gold-plated, oxygen-free copper sample holder was employed to connect the 10 mK stage and the sample box, ensuring efficient thermal conduction for maintaining the sample at cryogenic temperatures. Furthermore, an outer-surface-polished, gold-plated copper thermal shield cylinder was mounted around the entire 10 mK stage to enclose all devices and samples, thereby effectively reducing thermal radiation from higher-temperature components. Additionally, 12 GHz low-pass filters and infrared filters were implemented on both the input and output lines to further minimize high-frequency noise. 

Fig. 4(b) shows the magnitude of $S_{11}$ for A1 resonators at room temperature and 16 mK, respectively. The deeper dip curvature of $\mid$$S_{11}$$\mid$ at 16 mK manifests the higher \emph{Q}-factor compared with that at room temperature. However, due to a small kink near the resonance possibly originated from a weak spurious mode, it is difficult to employ the data fitting for obtaining the internal \emph{Q}-factor ($Q\textsubscript{i}$), which accounts for the intrinsic losses including electrode-related thermoelastic damping and ohmic loss, material-related acoustic loss, and anchor dissipation, etc. Therefore, we utilized the 3dB-\emph{Q} method to track the temperature dependence of energy dissipation, where $Q\textsubscript{3dB}$ is directly extracted from the full-width at half-maximum of the $\mid$$S_{11}$$\mid$ resonance dip, as shown in Fig. 4(c). The $Q\textsubscript{3dB}$ gradually increases from room temperature to 10 K, saturates to 190 down to the base temperature. There is a fourfold increase in $Q\textsubscript{3dB}$ compared with room temperature values, indicating smaller dissipation at low temperatures. Here, we note that the $Q\textsubscript{3dB}$ and $Q\textsubscript{i}$ of A1 and A3 resonators measured by the \emph{S}-parameters are somewhat smaller than $Q\textsubscript{s}$ in Fig. 3(c) and 3(d), due to the device-to-device variation on the wafer and the introduced loss from the bonding wires for the low-temperature measurements. 

The resonant frequency $f\textsubscript{r}$ slightly shifts from 4.83 GHz at 300 K to 4.796 GHz at 150 K, which is attributed to the intrinsic material properties of the piezoelectric LN. Below 150 K, $f\textsubscript{r}$ exhibits a small upturn and remains around 4.8 GHz from 40 K to the base temperature of 16 mK. The power dependence of $Q\textsubscript{3dB}$ for A1 resonators at 16 mK is displayed in the inset plot of Fig. 4(c), showing unchanged behaviors with increasing driven power which hints the two-level system loss is negligible or obscured by other energy loss channels in the present devices.  

\begin{figure}
  \centering
  \includegraphics[width=\linewidth]{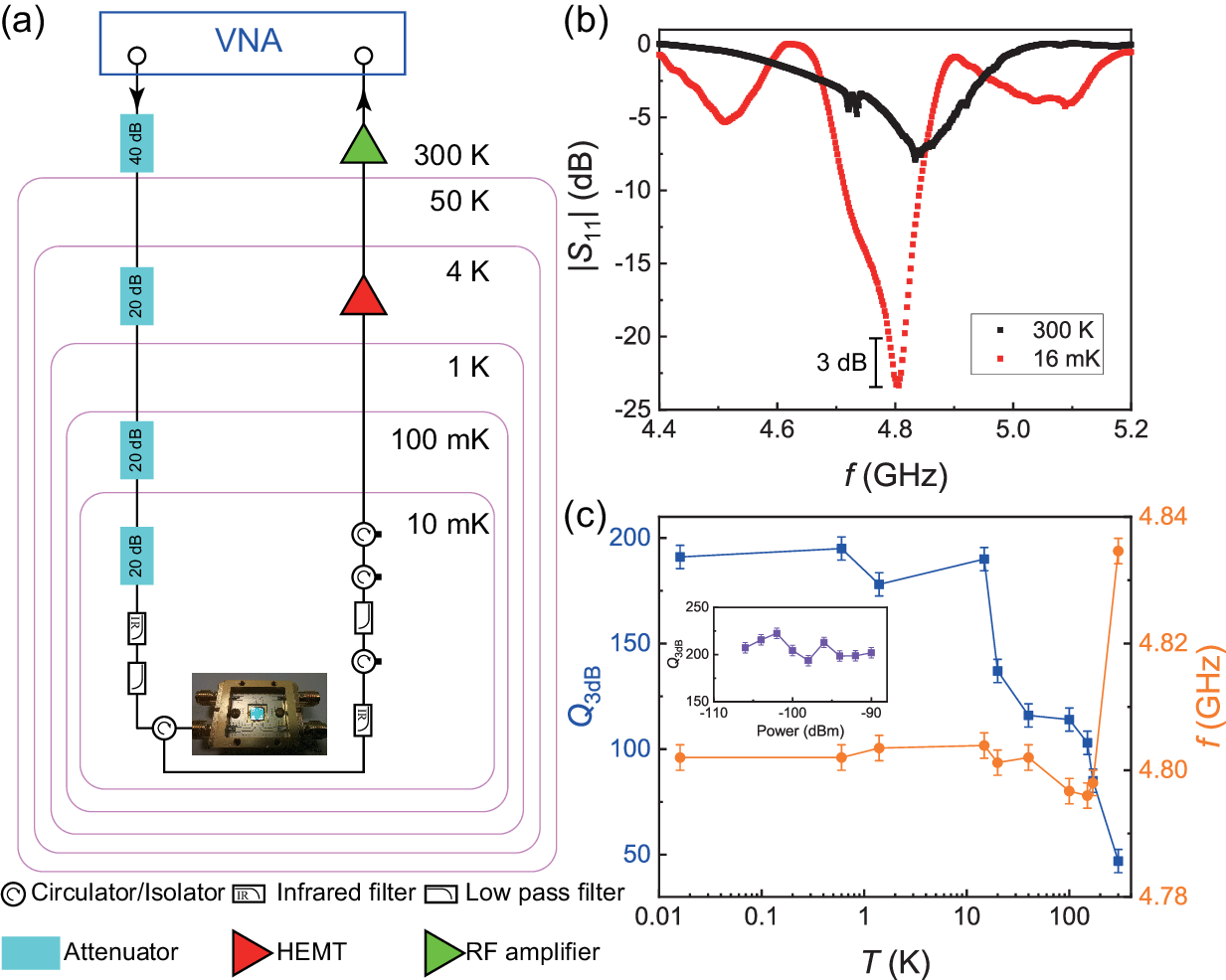}
  \caption{(a) Schematic diagram of low-temperature measurement setup for LWRs. (b) Measured magnitude of $S_{11}$ versus frequency for the A1 resonator at 300 K and 16 mK, respectively. (c) Extracted $Q\textsubscript{3dB}$ and resonant frequency $f\textsubscript{r}$ as a function of temperature. The inset plots $Q\textsubscript{3dB}$ as a function of on-chip input power at 16 mK. The error bars denote the variations from multiple measurements.}\label{Fig4}
\end{figure}

Figure 5 shows the low-temperature measurement results of A3 resonators. Compared with the A1 modes, the A3 mode is free from spurious modes, hence we fitted the magnitude and phase of the $S_{11}$ data at 12 mK using the formula
\begin{equation}
S_{11}(f) = \frac{(Q\textsubscript{c}-Q\textsubscript{i})/Q\textsubscript{c} + 2iQ\textsubscript{i}(f-f\textsubscript{r})/f}{(Q\textsubscript{c}+Q\textsubscript{i})/Q\textsubscript{c} + 2iQ\textsubscript{i}(f-f\textsubscript{r})/f},  \label{Eq2} 
\end{equation}
where \emph{Q}\textsubscript{c} denotes the coupling \emph{Q}-factor, representing the coupling rate with the external feedline circuits \cite{manenti2016surface}. Based on the fittings in Fig. 5(a) and 5(b), the $Q\textsubscript{i}$ and $Q\textsubscript{c}$ of the A3 resonator are obtained to be 105 and 180, respectively. For analytical efficiency, we adopted the Lorentz fitting to extract the temperature dependent $Q\textsubscript{i}$ hereafter, as this method yields results that are in agreement with the formula fitting method described above. Fig. 5(c) illustrates the normalized $\mid$$S_{11}$$\mid$ and the corresponding fittings across the superconducting transition temperature ($T\textsubscript{c} \simeq$ 1.2 K) of Al electrodes. Below $T\textsubscript{c}$, $\mid$$S_{11}$$\mid$ exhibits sharper resonances and the extracted $Q\textsubscript{i}$ undergoes a jump because of the negligible electrical loss in the superconducting state. Based on the value change of $Q\textsubscript{i}$ near $T\textsubscript{c}$, we estimated the internal loss has been reduced by about 23$\%$ due to the diminished electrode ohmic loss. Fig. 5(d) plots the extracted $Q\textsubscript{i}$ and $f\textsubscript{r}$ in the whole temperature range. As the temperature decreases from 300 K to 40 K, $Q\textsubscript{i}$ drops sharply, reaching a minimum value. Upon further cooling, $Q\textsubscript{i}$ begins to increase inversely and undergoes a sharp increment below $T\textsubscript{c}$ of Al electrodes. Meanwhile, the resonant frequency $f\textsubscript{r}$ increases linearly from 7.409 GHz to 7.431 GHz between 300 K and 100 K. Below 40 K, $f\textsubscript{r}$ decreases slightly and stabilizes around 7 GHz until the lowest temperature of 12 mK. It is worth noting that $Q\textsubscript{i}$ and $f\textsubscript{r}$ of A3 mode resonators could not be accurately determined between 40-100 K due to the degraded signal-to-noise ratio in this temperature range.

\begin{figure}
  \centering
  \includegraphics[width=\linewidth]{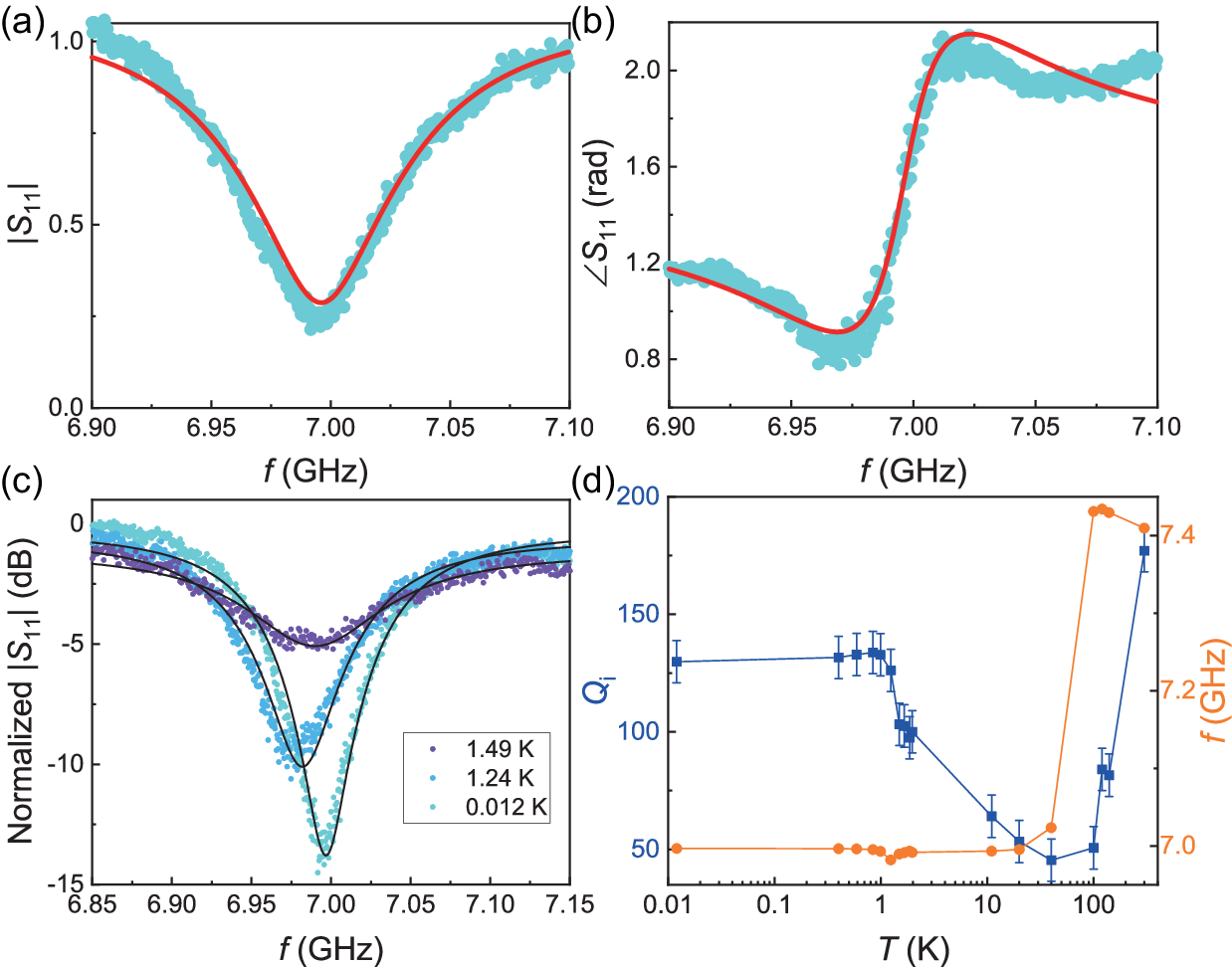}
  \caption{Magnitude (a) and phase (b) responses of the A3 resonator at 12 mK. The experimental and fitted reflection data $S_{11}$ are represented as symbols and solid lines, respectively. (c) The normalized $\mid$$S_{11}$$\mid$ at various temperatures across $T\textsubscript{c}$ of Al electrodes, along with the corresponding Lorentz fittings represented by the solid lines. (d) The temperature revolution of $Q\textsubscript{i}$ and $f\textsubscript{r}$ of the A3 resonator. The error bars denote the variations from multiple measurements and fitting errors.}\label{Fig5}
\end{figure}

Actually, the degradation of the A3 mode $Q\textsubscript{i}$ upon initially decreasing temperatures to 40 K resembles the low-temperature behavior observed in LN A1 mode acoustic filters and suspended wine-glass disk micromechanical resonators \cite{zheng2024temperature,li2009quality}. This phenomenon is attributed to the mechanical loss induced by the thin film deformation and phonon-phonon interactions. In contrast, the A1 mode $Q\textsubscript{i}$ gradually improves with cryogenic cooling due to the suppressed thermoelastic damping, consistent with the features of AlN-metal piezoelectric LWRs and SAW devices \cite{yamamoto2023low,tu2016effects,Wenzhen2025}. Nevertheless, the overall $Q\textsubscript{i}$ of our A1 mode devices is relatively small at low temperatures, indicating that the temperature-independent loss mechanisms dominate, such as anchor loss from supporting tethers and acoustic damping loss resulting from the surface damage during film thinning processes \cite{link2021a1}. Future optimizations of device design and fabrication, such as refining anchor geometry and enhancing film quality could be beneficial to improve the $Q\textsubscript{i}$ of LWRs.

In summary, we have systematically investigated the ultralow-temperature characteristics of high-frequency A1 and A3 mode resonators. The spurious modes of A1 resonators are successfully mitigated by utilizing the recessed electrode strategy, enabling effective coupling with superconducting qubits and high-performance RF filters. The \emph{Q}-factors of A1 modes demonstrate a fourfold enhancement at mK temperatures. Combined with the higher $k\textsubscript{t}^2$ of 9.79$\%$ compared to SAW resonators \cite{manenti2017circuit,bolgar2018quantum,ding2020enhanced}, the LN A1 mode resonators deliver a promising platform for strong-coupling cQAD. On the other hand, the A3 mode manifests elevated acoustic dissipation despite the diminished electrode-related ohmic loss in the superconducting state, revealing competing energy loss channels in these systems. These findings not only advance the development of higher order Lamb wave devices for quantum information processing but also highlight the significant potential of Lamb mode filters/sensors in future aerospace applications.

% If in two-column mode, this environment will change to single-column format so that long equations can be displayed. 
%\begin{widetext}
%$$\mbox{put long equation here}$$
%\end{widetext}

% Figures should be put into the text as floats. 
% Use the graphics or graphicx packages (distributed with LaTeX2e).
% Put the label that you will use with \ref{} command in the braces of the \label{} command.
%
% \begin{figure}
% \end{figure}

% \begin{table}
% \end{tabular}
% \end{table}

\begin{acknowledgments}
  We thank Ya Cheng and Zhiwei Fang for insightful discussions. This work was supported by the National Natural Science Foundation of China under Grant 62375274, Shanghai Technology Innovation Project under Grant XTCX-KJ-2023-01, and Key Research Program of the Chinese Academy of Sciences under Grant No. KGFZD-145-24-12. Jiazhen Pan acknowledges the support by the Natural Science Foundation of Jiangsu Province under Grant BK20230132. Hancong Sun acknowledges the support by the National Natural Science Foundation of China under Grant 62201396. The authors would like to thank ShanghaiTech Material and Device Lab (SMDL) for device fabrication.
\end{acknowledgments}

\section*{AUTHOR DECLARATIONS}
\section*{Conflict of Interest}
The authors have no conflicts to disclose.
\section*{Author Contributions}
Wenbing Jiang and Xuankai Xu contributed equally to this work.

\noindent \textbf{Wenbing Jiang}: Conceptualization (lead); Data curation (equal); Formal analysis (equal); Investigation (equal); Methodology (equal); Writing - review $\&$ editing (lead); Supervision (equal). \textbf{Xuankai Xu}: Conceptualization (equal); Data curation (equal); Formal analysis (equal); Investigation (equal). Writing - review $\&$ editing (equal). \textbf{Jiazhen Pan}: Data curation (equal); Formal analysis (equal); Investigation (equal). \textbf{Hancong Sun}: Data curation (supporting); Formal analysis (supporting); Investigation (supporting). \textbf{Yu Guo}: Data curation (supporting); Investigation (supporting). \textbf{Huabing Wang}: Supervision (supporting). \textbf{Libing Zhou}: Writing - review $\&$ editing (equal); Supervision (equal). \textbf{Tao Wu}: Conceptualization (equal); Formal analysis (equal); Methodology (equal); Writing - review $\&$ editing (equal); Supervision (lead).  
\section*{DATA AVAILABILITY}
The data that support the findings of this study are available from the corresponding author upon reasonable request.

\bibliography{Reference}

%\bibliography{maintext-bib}

\end{document}